\documentclass[11pt,a4paper]{article}

\usepackage{amsmath,amsthm,amsfonts}

\usepackage{cite}
\usepackage{mathptmx,times}

\newcommand{\HH}{\mathcal{H}}

\newcommand{\CC}{\mathbb{C}}
\newcommand{\RR}{\mathbb{R}}
\newcommand{\ZZ}{\mathbb{Z}}
\newcommand{\GG}{\mathcal{G}}

\newcommand{\dom}{\mathop{\mathrm{dom}}}
\newcommand{\tr}{\mathop{\mathrm{tr}}}
\newcommand{\indeg}{\mathop{\mathrm{indeg}}}
\newcommand{\outdeg}{\mathop{\mathrm{outdeg}}}

\newtheorem{theorem}{Theorem}
\newtheorem{prop}[theorem]{Proposition}
\newtheorem{lemma}[theorem]{Lemma}

\theoremstyle{definition}

\newtheorem{rem}[theorem]{\bf Remark}

\begin{document}

\title{\bf Spectra of  Schr\"odinger operators on equilateral quantum graphs}

\author{{\sc Konstantin Pankrashkin}\\[\bigskipamount]
Institut f\"ur Mathematik, Humboldt-Universit\"at zu Berlin,\\ Rudower Chaussee 25, 12489 Berlin, Germany\\
E-mail: const@mathematik.hu-berlin.de
}
\date{}

\maketitle

\begin{abstract}
We consider magnetic Schr\"odinger operators on quantum graphs
with identical edges. 
The spectral problem for the quantum graph is reduced to the discrete magnetic
Laplacian on the underlying combinatorial graph
and a certain Hill operator. In particular, it is shown that
the spectrum on the quantum graph
is the preimage of the combinatorial spectrum
under a certain entire function.
Using this correspondence we show that that the number
of gaps in the spectrum of the Schr\"odinger operators
admits an estimate from below in terms of the Hill
operator independently of the graph structure.
\end{abstract}

\section{Introduction}

Quantum graphs, also known as metric graphs or networks, are one-dimensional singular manifolds
together with self-adjoint differential operators on them.
Spectral and scattering properties of Schr\"odinger operators
in such structures attract a considerable interest during last years,
as they provide relevant models of nanostructures,
see e.g.~\cite{KS,KSM,KSmil,Ku1,Ku2} for a survey.
An important class of quantum graphs consists of graphs having
some regular structures; possible example may include trees
and periodic configurations, cf.~\cite{ms,Ku2,excl,bgp,bel}
and references therein.

It can be expected that there is a certain relationship
between a quantum graph and its combinatorial counterpart.
This question was first addressed at the physical level
of rigor in~\cite{alx} and the results are used intensively
in the physics literature for the study of network models
in the tight-binding approximation, see e.g.~\cite{oh}.
A mathematically rigorous correspondence between solutions
of continuous and combinatorial problems was given
in~\cite{exdual}, but the correspondence between the spectra
was possible only in very particular situations.

One may expect that the combinatorial structure of a quantum graph must
become especially visible in the case when the quantum graph is equilateral, i.e.
consists of identical segments with the same potentials on them.
This situation for ``free'' operators (i.e. without external potentials)
was actually considered already in~\cite{alx}; a mathematical justification
of the correspondence between the spectra for the discrete and continuous
Laplacian on a large class of graphs was given in~\cite{CC,bel}.
In both cases it appears that the spectrum on the quantum graph
essentially consists of the values $z$ such that $\cos\sqrt{z}$ lies in the spectrum
of the corresponding discrete (magnetic) Laplacian, which establishes
a useful correspondence between the quantum graphs and the well-developed theory
of combinatorial graphs, see e.g.~\cite{cdv,CDC,MW}.
On the other hand, the results obtained for the Laplacian only give
no information about the contribution
of the local metric properties and external potentials to the spectrum.
Up to now, spectra of quantum
graphs with  external potentials were studied only in rather particular cases,
see e.g.~\cite{RC,ms,mont,Ku2} and references therein.
We describe the spectrum of Schr\"odinger
operators on general equilateral graphs in the present paper.

We emphasize that the construction presented here differs essentially
from the constructions of~\cite{alx,CC,exdual} and is based on the Krein technique
of abstract self-adjoint boundary value problems~\cite{dm} similar to the one
used in the theory of explicitly solvable models~\cite{AGHH,bsp1}. This permits
us to minimize assumptions on the graph  and to obtain a rigorous relationship between different types of spectra.
For an arbitrary $L^2$ scalar potential on the edges, arbitrary
(not necessary identical) magnetic potentials and $\delta$-type boundary conditions
at the vertices we show that, up to a discrete set (the Dirichlet spectrum of the edges),
the spectrum a quantum graph $\Gamma$ with identical edges
has the form $\eta^{-1}\big(\sigma(2\Delta_\Gamma)\big)$, where $\Delta_\Gamma$
is the discrete magnetic Laplacian on the combinatorial graph
and $\eta$ is the discriminant (Lyapunov function) of a certain
singular Hill operator associated with the edge potential (see section~\ref{sec4}).
This provides, in particular, a violation of the Bethe-Sommerfeld conjecture, as very general
periodic structures appear to have infinitely many gaps. Note that the spectrum
of the Hill operator has the form $\eta^{-1}\big([-2,2]\big)$, so our result
is a natural extension of this classical situation, as $[-2,2]$ is exactly
the spectrum of the doubled one-dimensional discrete Laplacian.
In other words, we are able to separate completely the combinatorial
and metric properties of quantum graphs with identical edges.
In this way we arrive at a new possibility to
create gaps in the spectrum by a suitable choice of the scalar potential on the edges, which works
independently of graph structure and of the magnetic parameters.
Similar effects induced by changing the local geometry of graph are usually referred
to as decorations \cite{AS,Ku2} or, in a more abstract framework, models with inner
structure~\cite{bsp1}, so the description of the spectrum
obtained looks like ``decoration by potentials''; an interesting difference is that
the creation of gaps by decoration is possible outside the edge Dirichlet eigenvalues only~\cite{Ku2},
in contrast to the potential-induced gaps situated exactly around these eigenvalues.

\section{Schr\"odinger operators in  equilateral quantum graphs}\label{sec1}

Let $\Gamma$ be a countable  oriented graph.
The sets of the vertices and of the edges of $\Gamma$ will be denoted by $V$ and $E$, respectively.
We do not exclude multiple edges and self-loops.
For an edge $e\in E$ we denote by $\imath(e)$ its initial
vertex and by $\tau(e)$ its terminal vertex.
For a vertex $v$, the number of outgoing edges (outdegree) will be denoted
by $\outdeg v$ and the number of ingoing edges (indegree) will be denoted
by $\indeg v$. The degree of $v$ is $\deg v=\indeg v+\outdeg v$.
In what follows we assume that the degrees of the vertices of $\Gamma$
are bounded, $1\le\deg v\le N$ for all $v\in V$, in particular,
there are no isolated vertices. Note that each self-loop at $v$
counts in both $\indeg v$ and $\outdeg v$.

We introduce the discrete Hilbert space $l^2(\Gamma)$
consisting of functions on $V$ which
are summable with respect to the weighted scalar product,
$\langle f,g\rangle_\Gamma=\sum_{v\in V} \deg v \,\overline{f(v)}g(v)$.
As a set, this space coincides with the usual space of square integrable
functions on $V$ due to our restrictions on the degrees of the vertices,
but has a different Hilbert structure, and we prefer
to write $l^2(\Gamma)$ instead of $l^2(V)$
and to use the subindex $\Gamma$ in the scalar product
to avoid confusing.

To each edge $e\in E$ we assign a real number $\beta(e)$ (magnetic flux).
One can associate with this function $\beta$ a discrete magnetic Laplacian in $l^2(\Gamma)$,
\begin{equation}
        \label{eq-qmz}
\Delta_\Gamma h(v)=\frac{1}{\deg v}\Big(\sum_{e:\imath(e)=v} e^{-i\beta(e)}h\big(\tau(e)\big)+
\sum_{e:\tau(e)=v} e^{i\beta(e)}h\big(\imath(e)\big)
\Big).
\end{equation}
This expression defines a self-adjoint operator in $l^2(\Gamma)$, and for each $h\in l^2(\Gamma)$
there holds
\begin{multline*}
\|\Delta_\Gamma h\|^2_\Gamma=\sum_{v\in V}\frac{1}{\deg v}
\Big| \sum_{e:\imath(e)=v} e^{-i\beta(e)}h\big(\tau(e)\big)+
\sum_{e:\tau(e)=v} e^{i\beta(e)}h\big(\imath(e)\big)\Big|^2\\\le
\sum_{v\in V} \Big(
\sum_{e:\imath(e)=v} \big|h\big(\tau(e)\big)\big|^2
+
\sum_{e:\tau(e)=v} \big|h\big(\imath(e)\big)\big|^2
\Big)\\
=\sum_{w\in V} \deg w\,\big|h(w)\big|^2=\|h\|^2_\Gamma,
\end{multline*}
i.e. $\|\Delta_\Gamma\|\le 1$.

Now let us start with the construction of the continuous version of $\Gamma$.
Each edge $e$ of $\Gamma$ will be identified with segment $[0,1]$, such that
$0$ is identified with the vertex $\imath(e)$ and $1$ is identified with the vertex
$\tau(e)$. 
The topological space obtained is usually called a metric graph associated with $\Gamma$.

Magnetic Schr\"odinger operator in such a structure can be defined following the standard
construction of~\cite{Ku1,KSM}. 
The phase space of the graph is $\HH=\bigoplus_{e\in E}\HH_e$,
$\HH_e=L^2[0,1]$, consisting of functions $f=(f_e)$, $f_e\in \HH_e$.
On each edge consider the same scalar potential $U\in L^2[0,1]$.
We will distinguish between two types of graphs:
\begin{itemize}
\item[(A)] Even graphs, in which $\indeg v=\outdeg v$ for all $v\in V$; in this case there are no additional assumptions on $U$.
\item[(B)] All other graphs which do not satisfy (A); here we assume that $U$ is even, $U(x)=U(1-x)$.
\end{itemize}
Let $a_e\in C^1[0,1]$ be magnetic potentials on the edges, $e\in E$.
Associate with each edge a differential
expression $L_e:=(i\partial +a_e)^2+U$. The maximal operator which can be associated with these differential
expressions acts as
$(g_e)\mapsto(L_e g_e)$ on functions $g\in\bigoplus H^2[0,1]$. The integration by parts shows that this operator
is not symmetric, and it is necessary to introduce boundary conditions at the vertices to obtain
a self-adjoint operator. The standard self-adjoint boundary
conditions for magnetic operators are
\begin{gather*}
g_e(1)=g_b(0)=:g(v)\quad \text{ for all } b,e\in E \text{ with } \imath(b)=\tau(e)=v,\\
\sum_{e:\imath(e)=v} \big(g'_e(0)-ia_e(0)g_e(0)\big)-\sum_{e:\tau(e)=v} \big(g'_e(1)-ia_e(1)g_e(1)\big)=\alpha(v) g(v),
\end{gather*}
where $\alpha(v)$ are real numbers, the so-called coupling constants. The case
$\alpha(v)=0$ is called the Kirchhoff coupling at $v$.
If $a_e\equiv0$, this case appears, for example,
when graphs are considered as the limits of collapsing manifolds~\cite{EP}.

The gauge transformation $g_e(t)=\exp\Big(i\displaystyle\int_0^ta_e(s)ds\Big)f_e(t)$ removes the magnetic
potentials from the differential expressions, $\big((i\partial +a_e)^2+U\big)g_e=-f''_e+Uf_e$,
but the magnetic field
enters the boundary conditions through the parameters $\beta(e)=\displaystyle\int_0^1 a_e(s)\,ds$ in the following way:
\begin{subequations}
\begin{gather}
  \label{eq-fv1}
e^{i\beta(e)}f_e(1)=f_b(0)=:f(v),\quad \text{ for all } b,e\in E \text{ with } \imath(b)=\tau(e)=v,\\
  \label{eq-fv2}
f'(v):=\sum_{e:\imath(e)=v} f'_e(0)-\sum_{e:\tau(e)=v} e^{i\beta(e)}f'_e(1)=\alpha(v) f(v).
\end{gather}
\end{subequations}
The self-adjoint operator on $\HH$ acting as $(f_e)\mapsto (-f''_e+Uf_e)$ on functions
$(f_e)\in\bigoplus H^2[0,1]$ satisfying the boundary conditions~\eqref{eq-fv1}
and~\eqref{eq-fv2} for all $v\in V$
will be denoted by $\Lambda$. This is our central object.
Our aim is to find a relationship between the properties of the continuous operator
$\Lambda$ in $\HH$ and of the discrete operator \eqref{eq-qmz} in $l^2(\Gamma)$.
 
\section{Boundary triples and Krein's resolvent formula}

To deal with operators on graphs we describe very briefly some constructions
from the abstract theory of boundary value problems; for a detailed discussion we refer to~\cite{dm,gg}.

Let $S$ be a closed linear operator in a Hilbert space $\HH$ with the domain $\dom S$.
Assume that there exist
an auxiliary Hilbert space $\GG$ and two linear maps $\pi,\pi':\dom S\to \GG$ such that
\begin{itemize}
\item for any $f,g\in\dom S$ there holds
$\langle f,Sg\rangle-\langle Sf,g\rangle=\langle  \pi f,\pi' g\rangle-\langle\pi'f,\pi g\rangle$,
\item the map $(\pi,\pi'):\dom S\to\GG\oplus\GG$ is surjective,
\item the set $\ker\,(\pi,\pi')$ is dense in $\HH$.
\end{itemize}
The triple $(\GG,\pi, \pi')$ with the above properties
is called a \emph{boundary triple} for $S$.
If an operator $S$ has a boundary triple, then it has self-adjoint restrictions. For example,
if $A$ is a self-adjoint operator in $\GG$, then the restriction of $S$ to elements
$f$ satisfying abstract boundary conditions $\pi' f=A\pi f$ is a self-adjoint operator in $\HH$, which we denote by $H_A$.
Another example is the ``distinguished'' restriction $H$ corresponding to the boundary conditions $\pi f=0$.
The resolvents of  $H$ and $H_A$ as well as their spectral properties are connected by
Krein's resolvent formula, which will be described now.

Let $z\notin\sigma(H)$. For $g \in \GG$ denote by $\gamma(z) g$ the unique solution to
the abstract boundary value problem
$(S-z)f=0$ with $\pi f=g$ (such a solution exists due to the above conditions for $\pi$
and $\pi'$). Clearly, $\gamma(z)$ is a linear map from $\GG$ to $\HH$ and is an isomorphism between
$\GG$ and $\ker(S-z)$.
Denote also by $M(z)$ the bounded linear operator on $\GG$ given by $M(z)g=\pi' \gamma(z)g$; this operator
will be referred
to as the Weyl function (or the abstract Dirichlet-to-Neumann map) corresponding to the boundary triple $(\GG,\pi,\pi')$. The operator-valued functions $\gamma$ and $M$ are analytic outside $\sigma(H)$,
and $M(z)$ is self-adjoint for real $z$.
If all these maps are known, one can relate the operators
$H_A$ and $H$ as follows, cf. proposition 2 in~\cite{dm}.
\begin{prop}[Krein's resolvent formula]\label{krein} For $z\notin \sigma(H)\cup\sigma(H_A)$ the operator $M(z)-A$ acting on $\GG$ has a
bounded inverse defined everywhere, and $(H-z)^{-1}-(H_A-z)^{-1}=\gamma(z)\big(M(z)-A\big)^{-1}\gamma^*(\Bar z)$.
In particular, the set $\sigma(H_A)\setminus\sigma(H)$ consists exactly of
real numbers $z$ such that $0\in\sigma\big(M(z)-A\big)$.
The same correspondence holds for the point spectra, and $\gamma$ is an isomorphism
of the corresponding eigensubspaces.
\end{prop}

\section{Dirichlet-to-Neumann maps and Hill discriminants}\label{sec1d}

Let us describe an important example of a boundary triple and some related constructions
which will be actively used below.
Let $U\in L^2[0,1]$ be the potential on the graph edges introduced above.
In $\HH=L^2[0,1]$
consider the operator $S=-\dfrac{d^2}{dt^2}+U$ with $\dom S=H^2[0,1]$,
then one can take $\GG=\CC^2$, $\pi f=\big(f(0),f(1)\big)$,
$\pi' f=\big(f'(0),-f'(1)\big)$. Let us calculate the corresponding
Weyl function ${m}(z)$.
The distinguished restriction $D$ of $S$ given by $\pi f=0$ is exactly
the Dirichlet realization,
$Df=-f''+Uf$, $\dom D=\{f\in H^2[0,l]:\,f(0)=f(1)=0\}$.
In what follows we denote the eigenvalues of $D$ by $\mu_k$, $k=0,1,2\dots,$ $\mu_0<\mu_1<\mu_2<\dots$.

Let two functions $s,c\in\ker(S-z)$ satisfy $s(0;z)=c'(0;z)=0$ and $s'(0;z)=c(0;z)=1$,
i.e. $s$ and $c$ are the fundamental solutions of $S$.
For their Wronskian one has $w(z)=s'(x;z)c(x;z)-s(x;z)c'(x;z)\equiv 1$,
and $s$, $c$ as well as their derivatives are entire functions of $z$.
For $z\notin\sigma(D)$ one has $s(1;z)\ne0$, and any function $f\in\ker (S-z)$ can be written as
\begin{equation}
            \label{eq-fxz}
f(x;z)=\frac{f(1)-f(0) c(1;z)}{s(1;z)}\,s(x;z)+f(0)c(x;z).
\end{equation}
The calculation of $f'(0)$ and $-f'(1)$ gives
\begin{equation}
               \label{eq-rsz}
\begin{pmatrix}
f'(0)\\-f'(1)
\end{pmatrix}
=m(z)
\begin{pmatrix}
f(0)\\f(1)
\end{pmatrix},
\quad
m(z)=
\frac{1}{s(1;z)}\,\begin{pmatrix}
-c(1;z) & 1\\
1 & -s'(1;z)
\end{pmatrix}.
\end{equation}
Therefore, in this case the Weyl function $m(z)$ is exactly the Dirichlet-to-Neumann map.
In can be directly seen that $m(z)$ is real and self-adjoint for real $z$ and has poles
at the Dirichlet eigenvalues $\mu_k$.

In what follows we need some properties of the function
$\eta(z;\alpha)=s'(1;z)+c(1;z)+\alpha s(1;z)$.
This function is closely connected with the operator
\begin{equation}
           \label{eq-kpp}
P_\alpha=-\dfrac{d^2}{dt^2}+U_p(t)+\alpha\sum_{n\in\ZZ}\delta(t-n),
\end{equation}
where $U_p$ is the periodic extension of $U$. The operator $P_\alpha$ is a self-adjoint
operator in $L^2(\RR)$ acting
as $f\mapsto -f''+U_pf$ on functions $f\in H^2(\RR\setminus\ZZ)$
satisfying the boundary conditions $f(n-)=f(n+)=:f(n)$, $f'(n+)-f'(n-)=\alpha f(n)$,
$n\in\ZZ$ (the operator can be also properly defined by means of quadratic forms).
In particular, $P_0$ is the usual Hill operator. Due to the presence of 
a periodic $\delta$-potential we refer to $P_\alpha$ as to a Kronig-Penney-type Hamiltonian.
Let $z\notin\sigma(D)$. By direct calculations it can be shown that for any function $f\in H^2_\text{loc}(\RR\setminus\ZZ)$ satisfying the above boundary conditions
and such that $-f''+U_pf=zf$ in $\RR\setminus\ZZ$
there holds
\[
\begin{pmatrix}
f'(1+)\\f(1+)
\end{pmatrix}=
T(z;\alpha)\begin{pmatrix}
f'(0+)\\f(0+)
\end{pmatrix}
\]
with
\[
T(z;\alpha)=\begin{pmatrix}
s'(1;z)+\alpha s(1;z) & c'(1;z)+\alpha c(1;z)\\
s(1;z) & c(1;z)
\end{pmatrix}.
\]

The function $\tr T(z;\alpha)\equiv\eta(z;\alpha)$ is called the discriminant
or the Lyapunov function of $P_\alpha$ and its properties are well studied, see e.g.~\cite{LS,HM,ssh}.
The spectrum of $P_\alpha$ consists exactly of real $z$
satisfying the inequality $|\eta(z;\alpha)|\le 2$.
For the Dirichlet eigenvalues $\mu_k$ one has $\eta(\mu_k;\alpha)\le -2$ for even $k$
and $\eta(\mu_k;\alpha)\ge 2$ for odd $k$.
The set $\eta^{-1}\big((-2,2)\big)$ is the union of  open intervals
$(a_k,b_k)$ with $a_k<b_k\le\mu_k\le a_{k+1} $ for all $k=0,1,\dots,$,
and the spectrum of $P_\alpha$ is the union of the segments $[a_k,b_k]$ called bands.
The derivative $\eta'$ does not vanish in the intervals $(a_k,b_k)$, and
$\eta$ is a homeomorphism between each band and the segment $[-2,2]$.
The intervals $(-\infty,a_0)$ and $(b_k,a_{k+1})$, if non-empty, are called gaps.
If for some $k$ there holds $b_k=\mu_k=a_{k+1}$, then
$(b_k,a_{k+1})=\oslash$, and one says that the gap near $\mu_k$ is closed;
otherwise, for $b_k<a_{k+1}$, one says that the gap near $\mu_k$
is open. If only finitely many  gaps of $P_0$ are open, the potential $U$
is called a finite band potential, otherwise an infinite gap potential.
A study of periodic problems
with even more general singular potentials can be found,
for example, in~\cite{EK}.

\section{Relationship between the spectra of Schr\"odinger operators and discrete Laplacians}\label{sec4}

To work with the operator $\Lambda$ using the Krein resolvent formula
it is necessary to find a suitable linear operator $\Pi\supset\Lambda$ 
whose Weyl function has the possibly simplest form.
We define such an operator as follows. Set
$\dom\Pi=\{f\in\bigoplus H^2[0,1]:\,\text{Eq.~\eqref{eq-fv1} holds}\}$
and $\Pi(f_e)=(-f''_e+Uf_e)$.

\begin{lemma}\label{prop-pgz}
The operator $\Pi$ is closed.
For $f\in\dom\Pi$ put
\[
\pi f=\big(f(v)\big)_{v\in V},\quad
\pi' f=\Big(\dfrac{f'(v)}{\deg v}\Big)_{v\in V},
\]
where $f(v)$ and $f'(v)$ are given by~\eqref{eq-fv1} and~\eqref{eq-fv2}. The triple
$\big(l^2(\Gamma),\pi,\pi'\big)$ is a boundary triple for $\Pi$.
\end{lemma}

\begin{proof}
Denote by $L$ the direct sum of the operators $S$ from section~\ref{sec1d} over all edges.
Clearly, this operator is closed. On the domain of $L$ consider linear functionals
$l_{v,b,e}(f):=e^{i\beta(e)}f_e(1)-f_b(0)$, $\imath(b)=\tau(e)=v$, $v\in V$.
These functionals are bounded with respect to the graph norm of $L$ due to the Sobolev embedding,
therefore, the restriction of $L$ to the null space of all these functionals is a closed
operator. This restriction
is exactly the operator $\Pi$.

The Sobolev embedding theorem shows that $\pi f,\pi' f\in l^2(\Gamma)$
for any $f\in\dom\Pi$. Let $f,g\in \dom\Pi$, then
\begin{multline*}
%\[
\langle f,\Pi g\rangle-\langle \Pi f,g\rangle=
\sum_{e\in E} \big(\langle f''_e-Uf_e,g_e\rangle-\langle f_e,g''_e-Ug_e\rangle\big)\\
{}=\sum_{e\in E}\big(\langle f''_e,g_e\rangle-\langle f_e,g''_e\rangle\big).
%\]
\end{multline*}
Applying the partial integration one obtains
\begin{multline*}
\sum_{e\in E}
\big(\langle f''_e,g_e\rangle-\langle f_e,g''_e\rangle\big)\\
{}=\sum_{e\in E}\Big(
\overline{f_e(0)}g'_e(0)-\overline{f'_e(0)}g_e(0)
{}+\overline{f'_e(1)}g_e(1)-\overline{f_e(1)}g'_e(1)\Big)\\
{}=\sum_{v\in V}\bigg[
\sum_{e:\imath(e)=v}\Big(\overline{f_e(0)}g'_e(0)-\overline{f'_e(0)}g_e(0)\Big)\\
{}+\sum_{e:\tau(e)=v}\Big(\overline{f'_e(1)}g_e(1)-\overline{f_e(1)}g'_e(1)\Big)
\bigg]\\
{}=\sum_{v\in V}\bigg[\sum_{e:\imath(e)=v}\Big(\overline{f(v)}\,g'_e(0)- \overline{f'_e(0)}g(v)\Big)\\
{}+\sum_{e:\tau(e)=v}\Big(\overline{f'_e(1)}e^{-i\beta(e)}g(v)-\overline{e^{-i\beta(e)}f(v)}g'_e(1)\Big)
\bigg],
\end{multline*}
and by regrouping the terms one arrives at
\begin{multline*}
\langle f,\Pi g\rangle-\langle \Pi f,g\rangle=\sum_{v\in V}\Big[ \overline{f(v)}\,g'(v)
-\overline{f'(v)}\,g(v)\Big]\\
=\sum_{v\in V}\deg v\Big[ \overline{f(v)}\,\dfrac{g'(v)}{\deg v}
-\overline{\dfrac{f'(v)}{\deg v}}\,g(v)\Big]=\langle \pi f,\pi'g\rangle_\Gamma-\langle \pi' f,\pi g\rangle_\Gamma.
\end{multline*}
To show that the map $(\pi,\pi'):\dom\Pi\to l^2(\Gamma)\oplus l^2(\Gamma)$
is surjective we take arbitrary $h,h'\in l^2(\Gamma)$ and
choose functions $f_{jk}\in H^2[0,1]$ with $f_{jk}^{(i)}(l)=\delta_{ij}\delta_{kl}$, $i,j,k,l\in\{0,1\}$.
For each $e\in E$ put
\[
f_e=h\big(\imath(e)\big)f_{00}+e^{-i\beta(e)}h\big(\tau(e)\big)f_{01}+h'\big(\imath(e)\big)f_{10}
-e^{-i\beta(e)}h'\big(\tau (e)\big)f_{11}
\]
and take $f=(f_e)$. Clearly, $f\in\bigoplus H^2[0,1]$, and at each vertex $v\in V$ one has
\begin{gather*}
f(v)=f_b(0)=e^{i\beta(e)}\,f_e(1)=h(v),\quad \imath(b)=\tau(e)=v,\\
f'(v):=\sum_{e:\imath(e)=v} h'(v)+\sum_{e:\tau(e)=v} h'(v)=\deg v\,h'(v).
\end{gather*}
The surjectivity is proved. The space $\ker(\pi,\pi')$ is dense in $\HH$
as it contains $\bigoplus C_0^\infty(0,1)$.
\end{proof}

The distinguished restriction $\Pi_0$ of $\Pi$ is the direct sum of the operators
$D$ over all edges. In particular, the spectrum of $\Pi_0$ coincides
with the spectrum of $D$. Let us calculate the corresponding maps $\gamma(z)$
and $M(z)$.

\begin{lemma}\label{prop-mm}
Let $z\notin\sigma(D)$. The map $\gamma(z)$ and the Weyl function $M(z)$ for the operator $\Pi$
and the boundary triple from lemma~\ref{prop-pgz} have the following form.
For each $h\in L^2(\Gamma)$ and each $e\in E$ one has
\begin{equation}
          \label{eq-gh}
\big(\gamma(z)h\big)_e(x)=\dfrac{\big(s(1;z)c(x;z)-s(x;z)c(1;z)\big)h\big(\imath(e)\big)+e^{-i\beta(e)}h\big(\tau(e)\big)}{s(1;z)},
\end{equation}
and the Weyl function is
\begin{equation}
            \label{eq-mz}
M(z) =\dfrac{1}{2}\big({m}_{11}(z)+{m}_{22}(z)\big)\,\mathrm{id}
+{m}_{12}(z)\Delta_\Gamma,
\end{equation}
where $\Delta_\Gamma$ is the discrete magnetic Laplacian~\eqref{eq-qmz}
and $m_{jk}$ are the entries of the matrix~\eqref{eq-rsz}.
\end{lemma}

\begin{proof}
Take arbitrary $h\in l^2(\Gamma)$ and let $f=(f_e)\in \ker(\Pi-z)$ with
$f(v)=h(v)$ for all $v\in V$. Then
$-f_e''+Uf_e=zf_e$  for all $e\in E$. The condition \eqref{eq-fv1} reads as
$f_e(0)=h\big(\imath(e)\big)$ and $f_e(1)=e^{-i\beta(e)}h\big(\tau(e)\big)$ for all $e\in E$,
and Eq.~\eqref{eq-fxz} takes the form~\eqref{eq-gh}.
Applying the Dirichlet-to-Neumann map ${m}$ from \eqref{eq-rsz} to each component $f_e$
one obtains
\begin{align*}
f'_e(0)&={m}_{11}(z)f_e(0)+{m}_{12}(z)f_e(1)\\&={m}_{11}(z)h\big(\imath(v)\big)+{m}_{12}(z)e^{-i\beta(e)}h\big(\tau(e)\big),\\
-f'_e(1)&={m}_{21}(z)f_e(0)+{m}_{22}(z)f_e(1)\\&={m}_{21}(z)h\big(\imath(e)\big)+{m}_{22}(z)e^{-i\beta(e)}h\big(\tau(e)\big),
\end{align*}
and
\begin{multline*}
\big(M(z)h\big)(v)=\dfrac{f'(v)}{\deg v}=\dfrac{1}{\deg v}\Big[\sum_{e:\imath(e)=v} f'_e(0)-\sum_{e:\tau(e)=v} e^{i\beta(e)}f'_e(1)\Big]\\
=\dfrac{1}{\deg v}\Big[\outdeg v\, {m}_{11}(z)h(v)+{m}_{12}(z)\sum_{e:\,\imath(e)=v} e^{-i\beta(e)}h\big(\tau(e)\big)\\
+ \indeg v\,{m}_{22}(z)h(v)+{m}_{21}(z)\sum_{e:\,\tau(e)=v}e^{i\beta(e)}h\big(\imath(e)\big)\Big].
\end{multline*}
For graphs of type (A) one has $\indeg v=\outdeg v=\dfrac{1}{2}\deg v$.
Taking into account the equality ${m}_{12}(z)\equiv {m}_{21}(z)$, see~\eqref{eq-rsz}, one arrives
at~\eqref{eq-mz}. For graphs of type (B) there holds ${m}_{11}(z)={m}_{22}(z)$, see for example
Lemma~3.1 in~\cite{RC}, and the same calculations work. 
\end{proof}

Until now we did not use any specific requirement for the coupling constants $\alpha(v)$
in~\eqref{eq-fv2}. In the rest of the paper
we assume that
\begin{equation}
      \label{eq-cca}
\alpha(v)=\dfrac{1}{2}\,\deg v\,\alpha,\quad \alpha\in\RR.
\end{equation}
The boundary condition~\eqref{eq-fv2} takes the form $\pi' f=\dfrac{\alpha}{2}\,\pi f$.

\begin{lemma}\label{prop-gamma}
The set $\Sigma'=\Big\{z\in\RR\setminus\sigma(D):\,\,0\in\sigma\big(M(z)-\dfrac{\alpha}{2}\,\mathrm{id}\big)\Big\}$
with $M(z)$ from~\eqref{eq-mz} coincides with $\eta^{-1}\big(\sigma(2\Delta_\Gamma)\big)\setminus\sigma(D)$ where
$\eta(z;\alpha)$
is the discriminant of the Kronig-Penney Hamiltonian $P_\alpha$ from~\eqref{eq-kpp}.
\end{lemma}

\begin{proof}
Substituting the explicit expressions for $m_{jk}$ from~\eqref{eq-rsz}
and noting that $m_{12}$ does not vanish, one rewrites
the condition $0\in\sigma\big(M(z)\big)$ as
$0\in\sigma\big(2\Delta_\Gamma-\eta(z;\alpha)\,\mathrm{id})$.
\end{proof}

The following theorem describes the relationship between the spectra
of $\Lambda$ and of $\Delta_\Gamma$; this is the main result of the work.
\begin{theorem}\label{thmain}
The spectrum of $\Lambda$
is the union  $\sigma(\Lambda)=\Sigma_0\cup\Sigma$, where
$\Sigma_0\subset\sigma(D)$ is a discrete set and $\Sigma=\eta^{-1}\big(\sigma(2\Delta_\Gamma)\big)$,
where $\eta(z;\alpha)$ is the discriminant of the Kronig-Penney Hamiltonian $P_\alpha$ from~\eqref{eq-kpp}.
Moreover, for $z\notin\sigma(D)$ the condition $z\in \sigma_i(\Lambda)$ is equivalent to $\eta(z)\in\sigma_i(2\Delta_\Gamma)$, $i\in\{\mathrm{p},\mathrm{pp},\mathrm{disc},\mathrm{ess}\}$,
and for $i=\mathrm{p}$ the map $\gamma$ from \eqref{eq-gh} is an isomorphism
of the corresponding eigensubspaces.
\end{theorem}

\begin{proof}
This is a direct consequence of proposition~\ref{krein}. The spectrum of $\Lambda$ outside $\sigma(\Pi_0)\equiv\sigma(D)$
is exactly the set $\Sigma'$ from lemma~\ref{prop-gamma} and has the requested form.
The correspondence between the point spectra $\sigma_\mathrm{p}(\Lambda)$
and $\sigma_\mathrm{p}(2\Delta_\Gamma)$ follows also from proposition~\ref{krein}.
Moreover, as $\eta$ is a homeomorphism, it preserves the convergence
and isolated points, and $\gamma$ preserves the multiplicity of eigenvalues.
Therefore, the same correspondence holds for $\sigma_\mathrm{pp}=\overline{\sigma_\mathrm{p}}$
and $\sigma_\mathrm{disc}$, and hence for $\sigma_\mathrm{ess}=\sigma\setminus\sigma_\mathrm{disc}$. 
\end{proof}

The analysis of the relationship between the spectrum of $D$
and the spectrum of the quantum graph in the general situation seems to be very difficult.
Nevertheless, some particular situations can be handled.
For example, if $\beta(e)\equiv0$ and the potential $U$ is even,
then each loop on $\Gamma$
supports an eigenfunction composed from the Dirichlet eigenfunctions on each edge
of the loop. A more detailed investigation for $U=0$ can be found e.g. in~\cite{CC}.
In some particular situations similar analysis is possible also for non-vanishing
$\beta(e)$, cf.~\cite{bgp}.

We emphasize that the operator $P_\alpha$ itself can be considered as the Hamiltonian
of a quantum graph whose vertices are elements of $\ZZ$ with edges between the nearest neighbors.
The corresponding discrete Laplacian is $\Delta_\Gamma h(n)=\dfrac{1}{2}\,\big(h(n-1)+h(n+1)\big)$
whose spectrum is $[-1,1]$, so $P_\alpha=\eta^{-1}\big([-2,2]\big)$ (as $P_\alpha$
has no eigenvalues). Theorem~\ref{thmain} can be viewed as a generalization
of the classical spectral analysis of the Hill equation to general graphs.

\section{Spectral gaps}

Due to the specific form of the discrete magnetic Laplacian
it is rather easy to understand the meaning of the potential $U$
and the coupling constant $\alpha$ from~\eqref{eq-cca}
in the creating of gaps in the spectrum
of $\Lambda$. 

\begin{theorem} \label{prop-gap}

\textup{(A)} Each gap of $P_\alpha$ is a gap of $\Lambda$, probably
with an eigenvalue of $D$ in it, and the number of gaps in the spectrum of $\Lambda$ is not less than
the number of gaps for $P_\alpha$. In particular, if $P_\alpha$
has infinitely many gaps, then the same holds for $\Lambda$.

\textup{(B)} The number of gaps of $\Lambda$ in each interval $(\mu_k,\mu_{k+1})$
coincides with the number of gaps for $\Delta_\Gamma$.
If $\sigma(\Delta_\Gamma)\ne [-1,1]$,
then the spectrum of $\Lambda$ has infinitely many gaps.

\textup{(C)} The spectrum of $\Lambda$ has band structure if and only if
the spectrum of $\Delta_\Gamma$ has band structure.

\textup{(D)} If $\Lambda$ has only a finite number of gaps, then the same holds for $P_\alpha$,
and $\sigma(\Delta_\Gamma)=[-1,1]$.
\end{theorem}

\begin{proof}
(A) We have emphasized before in section~\ref{sec1d} the equality
$\sigma(P_\alpha)=\eta^{-1}([-2,2])$. On the other hand,
due to the estimate $\|\Delta_\Gamma(\beta)\|_\Gamma\le 1$ mentioned above,
for the part $\Sigma$ of the spectrum of $\Lambda$
one has $\Sigma\equiv\eta^{-1}\big(\sigma(2\Delta_\Gamma)\big)\subset
\eta^{-1}([-2,2])=\sigma(P_\alpha)$ and $\RR\setminus\big(\sigma(P_\alpha)\cup\sigma(D)\big)\subset \RR\setminus\sigma(\Lambda)$.
The part $\Sigma_0$ is a discrete set and cannot overlap open gaps of $P_\alpha$.

(B) The function $\eta$ is a homeomorphism between each band $[a_k,b_k]$ of $P_\alpha$
and the segment $[-2,2]$ with $\eta(\{a_k,b_k\})=\{-2,2\}$.
Therefore, the sets $\sigma(\Delta_\Gamma)$ and $\sigma(\Lambda)\cap(a_k,b_k)\equiv
\eta^{-1}\big(\sigma(2\Delta_\Gamma)\big)\cap(a_k,b_k)$ have equal number of connected components.
Each connected component of $[-2,2]\setminus\sigma(2\Delta_\Gamma)$ produces a gap
of $\Lambda$ in each interval $(a_k,b_k)$.

The items (C) and (D) are obvious corollaries of (A) and (B).
\end{proof}

\begin{rem}
It is worth emphasizing that the above assertions do not make any specific
assumptions on the combinatorial structure of the graph $\Gamma$. Therefore,
it is possible to create gaps of $\Lambda$ near the Dirichlet eigenvalues of the edges
by changing the potential
$U$ on each edge and the coupling constant $\alpha$; the structure
of the graph can be ignored. For example,
if $U=0$ and $\alpha\ne0$, then all gaps of $P_\alpha$ are open, see chapter~III.2 in~\cite{AGHH}.
Another possibility
to have gaps near each $\mu_k$ is to take a convex $U$ with non-vanishing derivative~\cite{gs2}.
If $U$ is piecewise continuous and $\alpha\ne0$, there are infinitely many gaps too~\cite{gs}.
\end{rem}

\begin{rem}
Although the above construction works in a rather general situation, we emphasize
an important particular class of graphs, the so-called graphs of groups~\cite{MW}.
Let $G$ be a finitely generated group and let $(g_1,\dots,g_n)$, $g_j\in G$, $j=1,\dots,n$,
be a minimal generating system of $G$. Consider the graph $\Gamma(G)$ whose
vertices are the elements of $G$ and two vertices $v,w\in G$ are connected
by an edge from $v$ to $w$ iff $w=v g_j$ for some $j$ (if $g_k^{-1}=g_k$ for some $k$, then
the graph has multiple edges). The group $G$ acts on the graph by the
left translations $g:v\mapsto gv$.  The properties of the Laplacian
depend on $G$ and can be rather exotic, even the band spectrum
can be guaranteed only under additional assumptions like the positivity of the
Kadison constant~\cite{hig,my,sun}.
Nevertheless, theorem~\ref{prop-gap} holds and shows the location of gaps
also in this case.
\end{rem}

Consider the simplest case of $G=\ZZ^n$. As the set of generators take
the standard basis $g_j=(\delta_{jk})$, $k=1,\dots,n$.
The corresponding quantum graph
is the lattice $\ZZ^n\subset\RR^n$ whose nodes $v$ and $w$ are connected
iff $|v-w|=1$ and $v_j\le w_j$ for all $j\in\{1,\dots,n\}$.

\begin{prop}\label{prop-zz} Let $\Gamma=\Gamma(\ZZ^n)$, $n\ge 1$,
and $\beta(e)=0$ for all $e\in E$, $j=1,\dots,n$.
The spectrum of $\Lambda$ is the union $\sigma(P_\alpha)\cup\Sigma_n$,
where $\Sigma_1=\oslash$ and $\Sigma_n=\sigma(D)=\sigma_p(\Lambda)$ for $n\ge 2$.
The operator $\Lambda$ has infinitely many bands if and only if
the operator $P_\alpha$ has the same property.
\end{prop}

\begin{proof}
The discrete Laplacian in $\ZZ^d$ is
$
\Delta_\Gamma h(n)=\dfrac{1}{2n}\,\sum_{|m-n|=1} h(m).
$
The Bloch substitution $h(n)=e^{i\langle n,q\rangle}$, $q=(q_1,\dots,q_n)\in[-\pi,\pi)^n$ reduces
$\Delta_\Gamma$ to the multiplication with $\dfrac{1}{n}\big(
\cos q_1+\dots+\cos q_n\big)$; the spectrum is absolutely continuous
and covers the segment $[-1,1]$.
Therefore, the set $\eta^{-1}\big(\sigma(2\Delta_\Gamma)\big)$ is exactly the spectrum of $P_\alpha$.

If $n=1$, the operator $\Lambda$ is exactly $P_\alpha$ and has no eigenvalues.
Let $n\ge 2$, $\mu\in\sigma(D)$, and $ \phi$ be the corresponding eigenfunction of $D$
with $\phi'(0)=1$.
For each $v\in\ZZ^n$ consider a function $f$ whose only nonzero components
are $f_{(v,v+g_1)}=\phi$,  $f_{(v+g_1,v+g_1+g_2)}=\phi'(1)\phi$, $f_{(v,v+g_2)}=-\phi$,
and $f_{(v+g_2,v+g_1+g_2)}=-\phi'(1)\phi$, then $f$ is an eigenfunction
of $\Lambda$.
\end{proof}

The estimate obtained can be considered as a violation
of the Bethe-Sommerfeld conjecture if the $\ZZ^n$-lattice is considered
as a multidimensional system. For a large class of multidimensional Schr\"odinger
operators with magnetic fields in $\RR^n$ it is proved that there are only finitely many gaps in the spectrum, see e.g.
\cite{gmc,elt,skr} and references therein. In our case, if the operator $P_\alpha$ has infinitely many bands,
which is a generic property of one-dimensional operators,
then the same holds for the multidimensional lattice even without magnetic fluxes $\beta$.
Similar effects induced by geometric properties were studied in~\cite{bgl,excl}.

\section*{Acknowledgments} 
The work was supported by the Deutsche Forschungsgemeinschaft,
the Sonderforschungsbereich ``Raum, Zeit, Materie'' (SFB 647, Berlin), 
and the International Bureau of BMBF at the German Aerospace Center
(IB DLR, cooperation Germany -- New Zealand NZL 05/001).

\end{document}